\documentclass[pr, twocolumn, preprintnumbers, showpacs, footnoteadded, superscriptaddress,longbibliography]{revtex4-1}
\pdfoutput=1

\usepackage[T1]{fontenc}
\usepackage{amssymb}
\usepackage{graphicx}
\usepackage{amsmath}
\usepackage{hyperref}
\usepackage{tensor}
\usepackage{epstopdf}
\usepackage{extarrows}
\usepackage{graphicx,subfigure}

\newcommand{\td}{\mathrm{d}}
\newcommand{\e}{\mathrm{e}}

\def \R {{^{(2)}R}}
\def \pD {\mathfrak{D}}

\begin{document}
\title{\boldmath Cosmic censorship hypothesis and entropy bound on black holes in canonical ensemble}
\author{Run-Qiu Yang}
\email{aqiu@tju.edu.cn}
\affiliation{Center for Joint Quantum Studies and Department of Physics, School of Science, Tianjin University, Yaguan Road 135, Jinnan District, 300350 Tianjin, P.~R.~China}

\begin{abstract}
This paper argues that the weak cosmic censorship hypothesis implies that the Schwarzschild black hole has maximal entropy in all stationary black holes of fixing temperature, or equivalently, to store a same amount of information the Schwarzschild black hole has highest temperature. It then gives the independent mathematical proofs for 4-dimensional general static black holes and stationary-axisymmetric black holes which have ``$t$-$\phi$'' reflection isometry.  This result does not only provide a new universal bound between temperature and entropy of black holes but also offers us new evidence to support the weak cosmic censorship hypothesis.
 \end{abstract}
\maketitle

\section{Introduction}\label{intro}
In general relativity, the Penrose's and Hawking's theorems about singularity show that spacetime singularity will be inevitable if matters satisfy a few of very general conditions~\cite{Hawking:1994ss,Senovilla:2014gza}. To keep the predicability in general relativity, it has been conjectured for a long time that spacetime singularity arising from gravitational collapse of physically ``reasonable'' matter must be shrouded by an event horizon, which is called ``weak cosmic censorship'' hypothesis~\cite{Penrose2002,Hod:2008zza}. Despite its clear significance, however, a complete proof is still open~\cite{Krolak1986,Rangamani2005,Ong:2020xwv}. In the absence of a complete proof, theoretical tests of cosmic censorship are of significant value.

One of well-studied the theoretical test was proposed by Penrose at ealy 1970's. It considers an asymptotically flat solution of Einstein equations with matter satisfying the dominant energy condition. Then if a Cauchy slice of this solution contains an outer-trapped 2-surface $S$ of area $A(S)$, and if $M$ is the Arnowitt-Deser-Misner (ADM) mass of the data on the slice, the inequality $A(S)\leq 16\pi M^2$  must be true if the ``weak cosmic censorship'' is true. Though the proof in general case is still open, people have proven it in a large class of cases~\cite{Huisken2001,Bray2001,Bray:2003ns}. Particularly , if the initial data set just forms a stationary black hole, we have
%
\begin{equation}\label{penroseineq1}
  A_H\leq 16\pi M^2\,.
\end{equation}
Here $A_H$ is just the area of event horizon. The inequality~\eqref{penroseineq1} has been also generalized into asymptotically anti-de Sitter (AdS) black holes~\cite{Husain:2017cmj,Engelhardt:2019btp}. Recently, a modified version by taking quantum effects into account was also discussed in Refs.~\cite{Bousso:2019var,Bousso:2019bkg}.

We have known that the black hole is not only a mechanical system but also a thermodynamical system, of which the entropy is given by Bekenstein-Hawking entropy $S=A_H/4$ and temperature is given by Hawking temperature $T_H=\kappa/(2\pi)$, where $\kappa$ is the surface gravity of event horizon. It is clear that the Penrose inequality can be regarded as an entropy bound. As total energy is fixed, this is an entropy bound in microcanonical ensemble. Once we reconsider the inequality~\eqref{penroseineq1} from the thermodynamics, it is naturally to ask the question: what will happen if we consider the black hole in canonical ensemble? One natural expectation is that the Schwarzschild back hole may also have maximal entropy in canonical ensemble, i.e.
\begin{equation}\label{goal1}
  S\leq\frac1{16\pi T_H^2}~~~~\text{i.e.}~~A_H\leq\frac1{4\pi T_H^2}\,.
\end{equation}
This is not a trivial corollary of bound~\eqref{penroseineq1}, as the physics of black holes may be not equivalent in different ensembles. For example, under certain circumstances, the partition function obtained by using the path-integral approach turned out to depend on the boundary conditions~\cite{PhysRevD.50.6394,Comer1992,Quevedo:2013pba}. Thus, we cannot use the inequality~\eqref{penroseineq1} to directly argue that inequality~\eqref{goal1} must be true in canonical ensemble.

The inequality~\eqref{goal1} was first conjectured by Ref.~\cite{Visser:1992qh} according to the computation of Hawking temperature in static spherically symmetric black holes. However, to our knowledge, no any progress was achieved beyond the static spherically symmetric case up to now. This paper will make a first step towards the proof of inequality~\eqref{goal1} in general case. Particularly, we will argue that the bound~\eqref{goal1} is a necessary condition of ``weak cosmic censorship''.  Then we will prove that: in 4-dimensional Einstein's gravity theory, for a static black hole or a stationary-axisymmetric black hole which has the ``$t$-$\phi$'' reflection isometry~\cite{PhysRevLett.26.331}, if (i) weak energy condition is satisfied and (ii) horizon has topology $S^2\times R$, then inequality~\eqref{goal1} is always true.  The requirement (ii) is redundant in asymptotically flat black hole if we use dominant energy condition to replace (i)~\cite{hawking1972,Hawking:1973uf,robertwald1984,Galloway:2005mf}. Note that the temperature (i.e. surface gravity) is constant automatically in the case considered here~\cite{Racz:1995nh}.

\section{An physical heuristic argument}
\begin{figure}
  \begin{center}
\includegraphics[width=.3\textwidth]{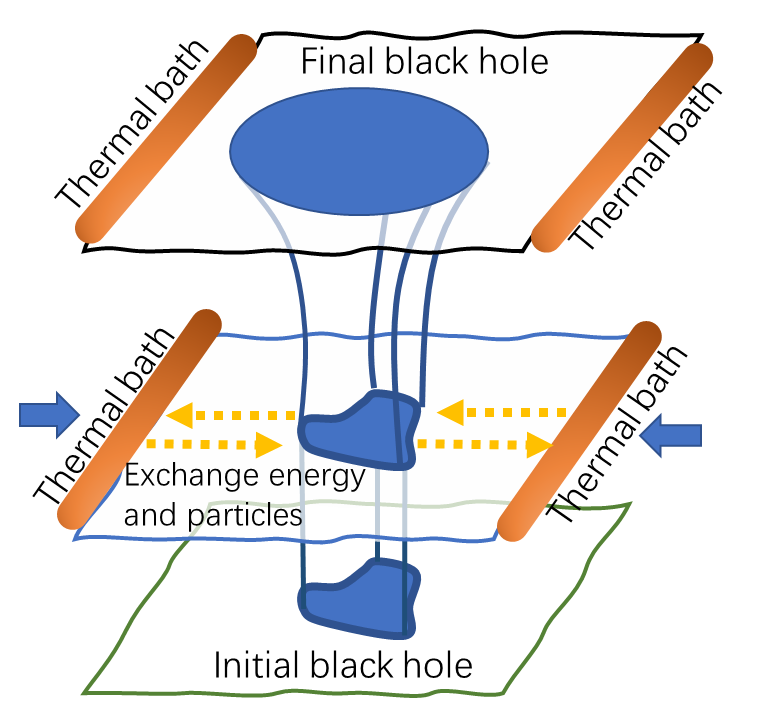}
  \caption{A schematic diagram of the ``thought experiment'' which shows how the inequality~\eqref{goal1} is relevant to weak cosmic censorship.}\label{schefig1}
  \end{center}
\end{figure}

Let us first argue that, similar to the Penrose inequality~\eqref{penroseineq1}, the bound~\eqref{goal1} is also implied by ``weak cosmic censorship''. Roughly speaking, the weak cosmic censorship states that the singularity originating from a gravitational collapse should be hidden by the event horizon. To connected this dynamics process with inequality~\eqref{goal1}, we consider following ``thought experiment'' shown by Fig.~\ref{schefig1}.
Firstly, we assume that there is a stationary initial black hole with temperature $T_H$ and horizon area $A_H$. Then we immersed it into a big thermal source at its boundary and the thermal source also has a fixed temperature $T_H$. Though the temperatures of thermal source and black hole are same, the black hole and thermal source may have different chemical potential. The black hole then will evolve by various isothermal processes and can exchange energy, particles, charges and so on with the thermal source.

It needs to note that, the vacuum black hole in asymptotically flat spacetime is unstable due to negative heat capacity. Thus, here we assume that exterior of black hole is full of classical matters which offer positive heat capacity during the evolution of the black hole and the dynamics of these isothermal processes is dominated by classical physics. The ``weak cosmic censorship'' guarantees that an asymptotically flat spacetime with regular initial conditions will be strongly asymptotically predictable~\cite{robertwald1984}. Then the null energy condition implies that the area of the event horizon will not decrease during these processes~\cite{Hawking1971}, i.e.
\begin{equation}\label{nodecresA}
  A_{H}\leq A_{H,f}\,.
\end{equation}
Here $A_{H,f}$ is the area of event horizon in final black hole. In physics, it is reasonable to expect the the black hole will settle down to a Kerr-Newman black hole by referring ``no-hair'' theorem of black hole. For the final Kerr-Newman black hole, let us assume that $M$ is the mass, $Ma$ is the angular momentum and $Q$ is the total charge. We then have following relationships
\begin{equation}\label{kerrA1}
  A_{H,f}=4\pi (r_h^2+a^2),~~T_H=\frac1{2\pi}\frac{r_h-M}{r_h^2+a^2}\,,
\end{equation}
with $r_h=M+\sqrt{M^2-a^2-Q^2}$.
We can verify that
\begin{equation}\label{kerrAT1}
T_H A_{H,f}=2r_h-2M\leq r_h\leq\sqrt{\frac{A_{H,f}}{4\pi}}\,.
\end{equation}
Combine Eqs.~\eqref{nodecresA} and \eqref{kerrAT1} and we will obtain desired inequality~\eqref{goal1}.

The reader will have noticed that the above argument makes a lot of global assumptions about the resulting space-times, and our current understanding is much too poor to be able to settle those one way or another. The ``no-hair'' theorem can also be broken in some physical acceptable situations~\cite{Hong:2019mcj,Hong:2020miv}. It is clear this heuristic cannot be treated as a valid proof. In following we will first give the proof for general static case and then discuss it in stationary-axisymmetric case.

\section{Coordinates gauge in Bondi-Sachs formalism}
In order to prove inequality~\eqref{goal1}, we will use Bondi-Sachs formalism, which foliates the spacetime by a series of null surfaces~\cite{BONDI1960,Sachs,Cao2013} and can be used for arbitrary spacetime. Here we first briefly explain about how to build such formalism when the black hole is static or stationary axisymmetric with ``$t$-$\phi$'' reflect isometry. In this paper, the Greek indexes $\{\mu, \nu,\cdots\}$ runs from 0 to 3 and the capital Latin indexes $\{A,B,\cdots\}$ run from 2 to 3.

\begin{figure}
  \begin{center}
\subfigure[]{\includegraphics[width=.25\textwidth]{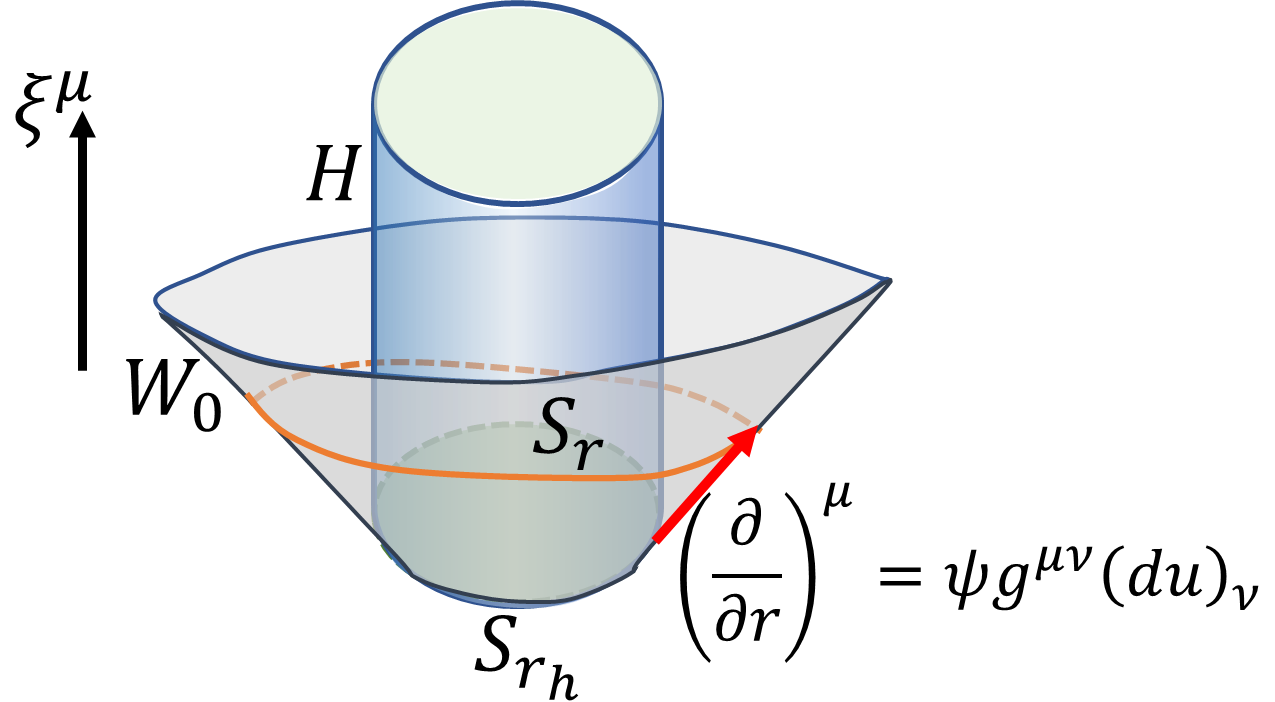}}
\subfigure[]{\includegraphics[width=.22\textwidth]{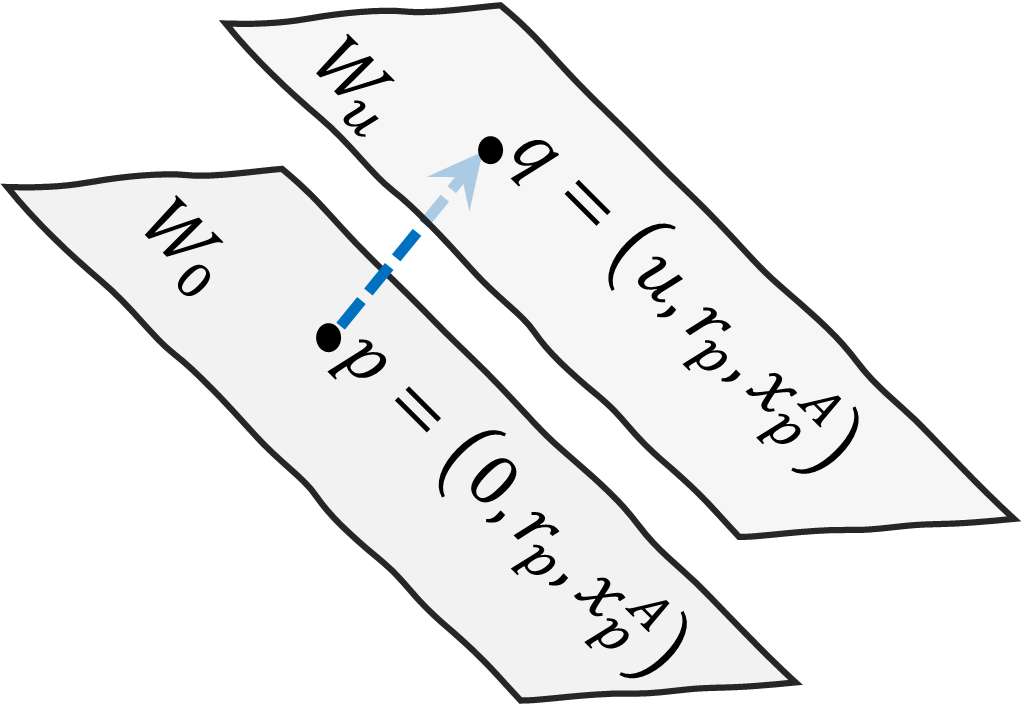}}
  \caption{(a) The schematic diagram about the event horizon $H$, spacelike cross-section $S_{r_h}$,  null hypersurface $W_0$ and equal-$r$ surface $S_r$. (b) For a point $q=\Phi_u(p)\in W_u$, we define its coordinate to be $(u, r_p, x_p^A)$, where $r_p$ and $x_p^A$ are the $r$ and $x^A$ coordinates of $p$. }\label{foliation1}
  \end{center}
\end{figure}

In static case or stationary axisymmetric case with ``$t$-$\phi$'' reflect isometry, the there is a Killing vector $\xi^\mu$ which is both tangent and normal to event horizon $H$. The Killing vector $\xi^\mu$ will generate a 1-parameter group of diffeomorphisms $\Phi_u$, i.e $\forall p$ in spacetime $\Phi_0(p)=p$ and the curve $\{\Phi_u(p)~|~u\in\mathbb{R}\}$ gives us an orbit of $\xi^\mu$.
Assume that $H$ is the event horizon (a 3-dimensional null surface) and $S_{r_h}$ is its one spacelike cross-section. As $H$ has topology of $S^2\times \mathbb{R}$, $S_{r_h}$ is a topology sphere The outward light rays of $S_{r_h}$ form a 3-dimensional null surface $W_0$, which is labeled by $u=0$. See Fig.~\ref{foliation1}(a).
Using map $\Phi_u$ we can obtain a series of equal-$u$ surfaces $W_u:=\Phi_u(W_0)$. We can prove that $W_u$ are all null by using the fact that $\xi^\mu$ is a Killing vector. These null surfaces are labeled by $u$=constant and so we have $g^{\mu\nu}(\td u)_\mu(\td u)_\nu=0$.

Let us now explain how to build $\{r,x^A\}$ coordinates. In the null surface $W_0$, the $r$-coordinate is choose to satisfy equation
\begin{equation}\label{defrcoord1}
  (\partial/\partial r)^\mu=\psi g^{\mu\nu}(\td u)_\nu
\end{equation}
with an arbitrary $u$-independent positive function $\psi$. The $r$-coordinate then is just the integral curve of Eq.~\eqref{defrcoord1}. We can adjust the zero point so that the $r$-coordinate of $S_{r_h}$ satisfies $r|_{S_{r_h}}=r_h$. The value of $r_h$ will be determined later. We denote equal-$r$ surface to be $S_r$ and introduce 2-dimensional coordinates $\{x^A\}$ in $S_r$.  The map $\Phi_u$ can bring $\{r,x^A\}$ from $W_0$ into all other null surfaces and so we obtain the coordinates $\{u,r,x^A\}$ for any point outside event horizon. As the result, we have $\xi^\mu=(\partial/\partial u)^\mu$. See Fig.~\ref{foliation1}(b) for a schematic explanation.

From Eq.~\eqref{defrcoord1} we see that $g_{rr}=g_{\mu\nu}(\partial/\partial r)^\mu(\partial/\partial r)^\nu=0$ and $g_{rA}=g_{\mu\nu}(\partial/\partial x^A)^\mu(\partial/\partial r)^\nu=\psi\partial u/\partial x^A$.  As the coordinate lines of $x^A$ lay in an equal-$u$ surface, we find $g_{rA}=\psi\partial u/\partial x^A=0$. Thus, the metric in coordinates $\{u,r,x^A\}$ has following form
\begin{equation}\label{BSmetric}
\begin{split}
  \td s^2&=-\frac{V}{r}\e^{2\beta}\td u^2-2\e^{2\beta}\td u\td r\\
  &+r^2h_{AB}(\td x^A-U^A\td u)(\td x^B-U^B\td u)\,.
  \end{split}
\end{equation}
At event horizon $H$ we have $V|_{H}=0$.
As the spacetime is asymptotically flat, we then fix the boundary conditions $\beta|_{r\rightarrow\infty}=0$, and
\begin{equation}\label{boundcs}
  \left.\frac{V}{r}\right|_{r\rightarrow\infty}=1,\quad h_{AB}|_{r\rightarrow\infty}\td x^A\td x^B=\td\hat{s}^2\,,
\end{equation}
Here $\td\hat{s}^2$ is the metric of unit sphere. In the asymptotic inertial frame, we require $r^2U^A\rightarrow0$ at the null infinity. It is also possible to choose a rotational frame with constant angular velocity, then we have $U^A=$constant. The $\psi$ in Eq.~\eqref{defrcoord1} is still free and we can fix this gauge freedom by a requirement
\begin{equation}\label{gaugeh1}
  \partial_rh=0\,.
\end{equation}
As the result, we have $\sqrt{h}\td^2x=\sqrt{h}|_{r\rightarrow\infty}\td^2 x=\td\Omega$ and $\td\Omega$ is the surface element of unit sphere. Then the area of event horizon has a simple formula
\begin{equation}\label{areaAh1}
  A_H=\int_{\text{fix}~u,~r=r_h}r^2\sqrt{h}\td^2x=4\pi r_h^2\,
\end{equation}
and we find $r_h=\sqrt{A_H/(4\pi)}$. The surface gravity of event horizon is given by
\begin{equation}\label{kappasqure}
  \kappa^2=-\left.\frac12(\nabla_\mu\xi_\nu)\nabla^\mu\xi^\nu\right|_H=-\frac18g^{\sigma\mu}g^{\tau\nu}(\td\xi)_{\sigma\tau}(\td\xi)_{\mu\nu}|_{H}\,.
\end{equation}
At event horizon $\kappa$ is a constant~\cite{hawking1972,Hawking:1973uf,robertwald1984,Galloway:2005mf}.
Strictly speaking, the metric~\eqref{BSmetric} with gauge condition~\eqref{gaugeh1} may cover only a neighborhood of event horizon. Here we assume that they have no coordinates singularity  in the whole spacetime outside horizon.

A proposition will be useful in our proofs: for any null vector $r^\mu$ which satisfies $\xi_\mu r^\mu<0$, if weak energy condition is satisfied, then $T_{\mu\nu}\xi^\mu r^\nu|_{H}\geq0$. Here $T_{\mu\nu}$ is the energy momentum tensor. The proof contains two steps. At the first step we use Raychaudhuri's equation at $H$
\begin{equation}\label{Raych}
\begin{split}
  \xi^\mu\partial_\mu\Theta&=\omega^{\mu\nu}\omega_{\mu\nu}-\kappa\Theta-\frac{\Theta^2}2-\sigma^{\mu\nu}\sigma_{\mu\nu}-R_{\mu\nu}\xi^\mu\xi^\nu\,.
  \end{split}
\end{equation}
where $\Theta$, $\kappa$, $\sigma_{\mu\nu}$ and $\omega_{\mu\nu}$ are the expansion, the surface gravity, the shear, and the rotation  of $\xi^\mu$, respectively. $R_{\mu\nu}$ is the Ricci tensor. As $\xi^\mu$ is a hypersurface-orthogonal null Killing vector at $H$, we find $\omega_{\mu\nu}=\sigma_{\mu\nu}=\Theta=0$. Then Einstein's equation and Eq.~\eqref{Raych} imply
\begin{equation}\label{Tuu0}
  T_{\mu\nu}\xi^\mu\xi^\nu|_{H}=R_{\mu\nu}\xi^\mu\xi^\nu|_{H}=0\,.
\end{equation}
In the second step, we take a vector $v^\mu=\xi^\mu+s r^\mu$. One can verify $v^\mu v_\mu|_{H}=2s\xi^\mu r_\mu$, so $v^\mu|_{H}$ is time-like for all $s>0$. The direct computation shows
\begin{equation}\label{Tvv1}
  T_{\mu\nu}v^\mu v^\nu|_{H}=2sT_{\mu\nu}\xi^\mu r^\nu|_{H}+s^2T_{\mu\nu}r^\mu r^\nu|_{H}
\end{equation}
The weak energy condition requires $T_{\mu\nu} r^\mu r^\nu\geq0$ and $T_{\mu\nu}v^\mu v^\nu|_{H}\geq0$ for all $s>0$. Then $T_{\mu\nu}\xi^\mu r^\nu|_{H}\geq0$ follows.

\section{Proof for static case}\label{static1}
We first focus on static case. We then choose the Killing vector $\xi^\mu=(\partial/\partial u)^\mu$ to be the one which stands for the static symmetry. So all components of metric is independent of $u$ and the reflection $u\rightarrow-u$ is a symmetry. Consider the induced metric of equal-$r$ surface,
\begin{equation}\label{BSmetric3}
  \td \tilde{s}_r^2=-\frac{V}{r}\e^{2\beta}\td u^2+r^2h_{AB}(\td x^A-U^A\td u)(\td x^B-U^B\td u)\,.
\end{equation}
The $\xi^\mu=(\partial/\partial u)^\mu$ lies in this subspacetime, so the reflection symmetry implies $U^A=0$.
Applying the metric~\eqref{BSmetric} with $U^A=0$, we find that Eq.~\eqref{kappasqure} reduces into
\begin{equation}\label{kappa1}
  \kappa^2=\left.-\e^{-4\beta}g_{uu}(\partial_r\sqrt{-g_{uu}})^2\right|_{H}\,.
\end{equation}
Using the fact $A_H=4\pi r_h^2$, we finally obtain
\begin{equation}\label{kappaAh1}
  \kappa\sqrt{A_H/4\pi}=\frac12\partial_rV|_H
\end{equation}

Now let us apply Einstein's equation. The Einstein's equation shows following two relevant equations~\cite{doi:10.1063/1.525796,M_dler_2016} (See appendix~\ref{hypereqs})
\begin{equation}\label{eqforbeta}
  \partial_r\beta=\frac{r}{16}h^{AC}h^{BD}(\partial_rh_{AB})(\partial_rh_{CD})+2\pi r T_{rr}
\end{equation}
and
\begin{equation}\label{eqforV}
\begin{split}
  \e^{-2\beta}\partial_rV&=\frac{\R}2-\pD^2\beta-(\pD\beta)^2\\
  &-8\pi r^2\e^{-2\beta}T_{ur}+4\pi rV\e^{-2\beta}T_{rr}\,.
  \end{split}
\end{equation}
Here $\R$ and $\pD_A$ stand for the scalar curvature and covariant derivative operator of $h_{AB}$.

Let us now prove $\partial_r\beta\geq0$ and $\beta\leq0$. As $T_{rr}$ is a ``null-null'' component of energy momentum tensor, the weak energy condition insures $T_{rr}\geq0$. 
We note that, for any surface of fixed $r$ and $u$ , the $\mathcal{X}_{AB}:=\partial_rh_{AB}$ is a tensor of the 2-dimensional space spanned by the coordinates $\{x^A\}$, i.e. $\mathcal{X}_{AB}$ will be transformed as a tensor under coordinates transformation $x^A\rightarrow\tilde{x}^A=\tilde{x}^A(x)$ (note that $\tilde{x}^A$ does not depend on $u$ and $r$). Then it is clear that $h^{AC}h^{BD}(\partial_rh_{AB})(\partial_rh_{CD})$ is invariant under such coordinates transformation. By using this invariance, we can always find a suitable coordinates transformation locally so that the inverse induced metric $h^{AB}$ has a diagonal form with two positive eigenvalues $\{\lambda^A\}$. In this special coordinates, the components of inverse induced metric become $\lambda^A\delta^{AB}$ (no summation) and we assume that the components of $\partial_rh_{AB}$ become $\tilde{\mathcal{X}}_{AB}$. Then we have
\begin{equation}\label{posithrh1}
\begin{split}
  &h^{AC}h^{BD}(\partial_rh_{AB})(\partial_rh_{CD})\\
  =&\sum_{A,B,C,D}\lambda^A\lambda^B\delta^{AC}\delta^{BD}\tilde{\mathcal{X}}_{AB}\tilde{\mathcal{X}}_{CD}\\
  =&\sum_{A,B}\lambda^A\lambda^B(\tilde{\mathcal{X}}_{AB})^2\geq0\,.
  \end{split}
\end{equation}
This proves $\partial_r\beta\geq0$ and so we find $\beta\leq\beta(\infty)=0$.

Take $r^\mu=(\partial/\partial r)^\mu$ and we will find that weak energy condition insures $T_{ur}|_{H}=T_{\mu\nu}\xi^\mu r^\nu|_{H}\geq0$. Then at horizon we have
\begin{equation}\label{eqforV2}
  \e^{-2\beta}\partial_rV|_{H}\leq\frac{\R}2-\pD^2\beta\,.
\end{equation}
Using the fact that $\partial_rV|_H\geq0$, $\beta\leq0$ and Eq.~\eqref{kappaAh1}, we find $\e^{-2\beta}\partial_rV|_H\geq\partial_rV|_H=2\kappa\sqrt{A_H/4\pi}$ and so
\begin{equation}\label{intgkappa1}
\begin{split}
  &\int_{H}\frac{\partial_rV}{\e^{2\beta}}\sqrt{h}\td^2x\geq2\kappa\sqrt{\frac{A_H}{4\pi}}\int_H\sqrt{h}\td^2x=2\kappa\sqrt{4\pi A_H}\,.
  \end{split}
\end{equation}
Here we have used the fact that $\kappa$ is constant at event horizon. Taking Eq.~\eqref{eqforV2} into Eq.~\eqref{intgkappa1}, we finally obtain
\begin{equation}\label{intgkappa2}
  2\kappa\sqrt{4\pi A_H}\leq\int_{r=r_h}\left[\frac{\R}2-\pD^2\beta\right]\sqrt{h}\td^2x=4\pi\,.
\end{equation}
Here we have neglected the totally divergent term and used Gauss-Bonnet theorem. Rewrite the surface gravity in terms of Hawking temperature $T_H=\kappa/(2\pi)$ and we will obtain the desired inequality~\eqref{goal1}.

\section{Proof of stationary-axisymmetric case with ``$t$-$\phi$'' reflect isometry}\label{tphisy1}

In stationary axisymmetric black hole, there are two commutative Killing vectors $t^\mu=(\partial/\partial t)^\mu$ (with $t^\mu t_\mu=-1$ at infinity) and $\Psi^\mu=(\partial/\partial\phi)^\mu$, which present the time translation symmetry and rotational symmetry respectively.  In addition, the Killing vector $\Psi^\mu$ is tangent to event horizon $H$ (but $t^\mu$ may not). In the study of stationary axisymmetric black holes, ``$t$-$\phi$'' reflected isometry is a usual assumption, which coves most of physical interesting cases~\cite{PhysRevLett.26.331,Wald:1999vt}. By this assumption, there is a constant $\Omega_H$ such that Killing vector $\xi^\mu=t^\mu+\Omega_H\Psi^\mu$ is both tangent and normal to event horizon. Here $\Omega_H$ is a constant and stands for the angular velocity of event horizon. The Hawking temperature then is given by the Killing vector $\xi^\mu$ rather than Killing vector $t^\mu$~\cite{Wald:1999vt}.

At horizon, we choose $S_{r_h}$ such that $(\partial/\partial\phi)^\mu$ lies on the surface $S_{r_h}$. Because of rotational symmetry, we can require the function $\psi$ in Eq.~\eqref{defrcoord1} to satisfy $\Psi^\mu\partial_\mu\psi=0$. As the result we can prove that the orbit of $\Psi^\mu$, i.e. the coordinate line of $\phi$, will always lie on a $S_{r}$. See appendix~\ref{app2} for a proof. We choose  coordinate $u$ by requiring $(\partial/\partial u)^\mu=\xi^\mu$ and choose coordinates $\{x^2=\theta, x^3=\phi\}$ for $S_r$.  At null infinity, this corresponds to a \emph{rotational frame} rather than an internal frame.

Though $U^A$ is not zero in general, it still satisfies a few of properties. Firstly, as the Killing vector $\xi^\mu$ is orthogonal to event horizon, we then have $\xi_\mu(\partial/\partial \phi)^\mu|_{H}=g_{\mu\nu}(\partial/\partial \phi)^\mu(\partial/\partial u)^\nu|_{H}=0$. This leads to $U^\phi|_{H}=0$. On the other hand, consider the induced metric of fixing $r$, i.e. metric~\eqref{BSmetric3}. The vector fields $(\partial/\partial u)^\mu$ and $(\partial/\partial\phi)^\mu$ are both tangent to equal-$r$ hypersurface, so vector $t^\mu$ is also tangent to equal-$r$ hypersurface. Thus, the orbits of two Killing vectors $t^\mu$ and $(\partial/\partial\phi)^\mu$ both lay in this time-like hypersurface. Then the ``$t$-$\phi$'' reflection isometry requires metric~\eqref{BSmetric3} is invariant under the transformation $\{t\rightarrow-t, \phi\rightarrow-\phi\}$. According to the relationship between $u, t$ and $\phi$, this requires that  metric~\eqref{BSmetric3} is invariant under the transformation $\{u\rightarrow-u, \phi\rightarrow-\phi\}$. This leads to $U^{\theta}=0$. To conclude, $U^A$ satisfies following two properties in our coordinates gauge,
\begin{equation}\label{uagauge1}
  U^{\phi}|_{H}=0,~~U^{\theta}=0\,.
\end{equation}
The formula of surface gravity is still given by Eq.~\eqref{kappasqure}. By using metric~\eqref{BSmetric} and Eq.~\eqref{uagauge1}, we finally find that the $\kappa$ is still give by Eq.~\eqref{kappaAh1}.

Now we apply Einstein's equation. The equation of $\beta$ is still given by Eq.~\eqref{eqforbeta} but Eq.~\eqref{eqforV} is replaced by~\cite{doi:10.1063/1.525796,M_dler_2016} (See appendix~\ref{hypereqs})
\begin{equation}\label{eqforV3}
\begin{split}
  &\e^{-2\beta}\partial_rV=\frac{\R}2-\pD^2\beta-(\pD\beta)^2+\pD_A\left[\frac{\e^{-2\beta}}{2r^2}\partial_r(r^4U^A)\right]\\
  &-\partial_r(r^4U^A)\pD_A\left(\frac{\e^{-2\beta}}{2r^2}\right)-\frac{8\pi r^2}{\e^{2\beta}}T_{ur}-8\pi r^2U^A\e^{-2\beta}T_{rA}\\
  &+4\pi rV\e^{-2\beta}T_{rr}-\frac{r^4}4\e^{-4\beta}h_{AB}(\partial_rU^A)(\partial_rU^B)\,.
  \end{split}
\end{equation}
Here all variables are independent of $\{u,\phi\}$. Note that $U^\theta=0$ leads to $[\partial_r(r^4U^A)]\pD_A(\e^{-2\beta}/r^2)=0$, so Eq.~\eqref{eqforV3} shows
\begin{equation}\label{eqforV4}
\begin{split}
  \left.\frac{\partial_rV}{\e^{2\beta}}\right|_{H}\leq\frac{\R}2-\pD^2\beta+\pD_A\left[\frac{\e^{-2\beta}}{2r}\partial_r(r^4U^A)\right]\,.
  \end{split}
\end{equation}
Here we also used the fact that $U^A|_{H}=0$, which results from Eq.~\eqref{uagauge1}.
Similar to Eq.~\eqref{intgkappa2}, after integrating Eq.~\eqref{eqforV4}, we can still obtain $2\kappa\sqrt{4\pi A_H}\leq 4\pi$ and so the bound~\eqref{goal1} follows.


\section{Conclusion and discussion}
To conclude, this paper proposes a new entropy bound~\eqref{goal1} for black holes in canonical ensemble and shows the Schwarzschild black hole has the maximal entropy. This is a parallel version Penrose inequality in the canonical ensemble. We argue that, in a certain circumstance, the bound~\eqref{goal1} will be necessary condition of ``weak cosmic censorship''. We then prove it in 4-dimensional general static case and stationary-axisymmetric case with the ``$t$-$\phi$'' reflection isometry. The bound~\eqref{goal1} also has an inverse interpretation: to store same a amount of information, the Schwarzschild black hole will have highest temperature.

It is interesting to study the generalization of bound~\eqref{goal1} in higher dimensional case, where we assume that the horizon has topology $S^{d-1}\times R$. The bound~\eqref{goal1} should become
\begin{equation}\label{goald1}
  A_H\leq\left(\frac{d-2}{4\pi T_H}\right)^{d-1}\Omega_{d-1}
\end{equation}
Here $\Omega_{d-1}$ is the surface area of $(d-1)$-dimensional unit sphere. To derive bound~\eqref{goald1} one sufficient condition is
\begin{equation}\label{yamabe1}
  \Omega_{d-1}^{-1}\int_{S_{r_h}} {^{(d-1)}R}\leq(d-1)(d-2)\,.
\end{equation}
Here ${^{(d-1)}R}$ is the scalar curvature of metric $h_{AB}$. In the case $d=3$, this follows from Gauss-Bonnet theorem. For the case $d\geq4$, the situation is less clear. Inequality~\eqref{yamabe1} will be true if the horizon has $(d-1)$-dimensional spherical symmetry. If the metric does not have this symmetry however we have, without further information, little control of the integrand. Note that the bound~\eqref{goald1} may still be true even if Eq.~\eqref{yamabe1} is broken.

This paper focuses on Einstein gravity theory. It is interesting to study the generalizations in other gravity theories in the future, such as coupling with a dilation field or adding higher order curvature terms. In our proof, we apply a special coordinates gauge to simplify the discussion. However, outside the event horizon, we cannot prove that such coordinates system exist globally in all cases. This leaves an issue for the further study. It is also worthy of studying how to prove bound~\eqref{goal1} by a coordinate-independent method.

\begin{acknowledgments}
This work is supported by the National Natural Science Foundation of China under Grant No. 12005155.
\end{acknowledgments}
\appendix
\section{Hypersurface equations in Bondi-Sachs formalism}\label{hypereqs}
In this appendix, we give a few of mathematical formulas in Bondi-Sachs formalism. In the main text, we focus on the static spacetime or axisymmetric stationary spacetime with $t$-$\phi$ reflection isometry. In fact, we can build local coordinates $\{u,r,x^A\}$ for arbitrary spacetime so that the metric has following form~\cite{BONDI1960,Sachs,Cao2013}
\begin{equation}\label{BSmetrica}
\begin{split}
  \td s^2&=-\frac{V}{r}\e^{2\beta}\td u^2-2\e^{2\beta}\td u\td r\\
  &+r^2h_{AB}(\td x^A-U^A\td u)(\td x^B-U^B\td u)\,.
  \end{split}
\end{equation}
and satisfies the gauge $\partial_rh=0$.  Here all components may depend on $\{u,r,x^A\}$. The corresponding non-zero components of the inverse metric are
\begin{equation}\label{inveresg}
\begin{split}
  &g^{ur}=-\e^{-2\beta},~~g^{rr}=\frac{V\e^{-2\beta}}r,\\
  &g^{rA}=-U^A\e^{-2\beta},~~g^{AB}=\frac{h_{AB}}{r^2}\,.
  \end{split}
\end{equation}
The two relative equations in the main text come from the null hypersurface constraint equations. Following Ref.~\cite{doi:10.1063/1.525796} and taking $\lambda=1$ (or following Ref.~\cite{M_dler_2016}), we will have
\begin{equation}\label{hyperbeta}
  \partial_r\beta=\frac{r}{16}h^{AC}h^{BD}(\partial_rh_{AB})(\partial_rh_{CD})+2\pi r T_{rr}
\end{equation}
and
\begin{equation}\label{hyperV}
\begin{split}
  &\e^{-2\beta}\partial_rV=\frac{\R}2-\pD^2\beta-(\pD\beta)^2\\
  &-\frac{r^4}4\e^{-4\beta}h_{AB}(\partial_rU^A)(\partial_rU^B)+\frac{\e^{-2\beta}}{2r^2}\pD_A[\partial_r(r^4U^A)]\\
  &+4\pi (r^2 T-h^{AB}T_{AB})\,.
  \end{split}
\end{equation}
Here $T_{AB}$ is the projection of energy momentum tensor on the subspace spanned by $\{x^A\}$ and $T=g^{\mu\nu}T_{\mu\nu}$.
Note here we use signature $(-,+,+,+)$ but Ref.~\cite{doi:10.1063/1.525796} used the signature $(+,-,-,-)$. The convention on the definition of Riemannian curvature is also different from Ref.~\cite{doi:10.1063/1.525796}. In the convention of Ref.~\cite{doi:10.1063/1.525796}, the unit sphere has curvature $\R=-2$, however, in this paper the unit sphere has curvature $\R=2$.

To obtain the equations used in the main text, we first note the facts
\begin{equation}\label{duA1}
\begin{split}
  &\frac{\e^{-2\beta}}{2r^2}\pD_A[\partial_r(r^4U^A)]\\
  =&\pD_A\left[\frac{\e^{-2\beta}}{2r^2}\partial_r(r^4U^A)\right]-\partial_r(r^4U^A)\pD_A\left(\frac{\e^{-2\beta}}{2r^2}\right)
  \end{split}
\end{equation}
and
\begin{equation}\label{traceT}
  \begin{split}
  T=&g^{uu}T_{uu}+2g^{ur}T_{ur}+2g^{uA}T_{uA}\\
  &+2g^{rA}T_{rA}+g^{rr}T_{rr}+r^{-2}h^{AB}T_{AB}
  \end{split}
\end{equation}
Using Eq.~\eqref{inveresg}, we find Eq.~\eqref{traceT} becomes
\begin{equation}\label{traceT2}
  \begin{split}
  T=&-2\e^{-2\beta}T_{ur}-2U^A\e^{-2\beta}T_{rA}+\frac{V\e^{-2\beta}}rT_{rr}\\
  &+r^{-2}h^{AB}T_{AB}\,.
  \end{split}
\end{equation}
Then we obtain
\begin{equation}\label{eqforV3b}
\begin{split}
  &\e^{-2\beta}\partial_rV=\frac{\R}2-\pD^2\beta+\pD_A\left[\frac{\e^{-2\beta}}{2r^2}\partial_r(r^4U^A)\right]\\
  &-(\pD\beta)^2-\frac14r^4\e^{-4\beta}h_{AB}(\partial_rU^A)(\partial_rU^B)\\
  &+4\pi rV\e^{-2\beta}T_{rr}-\partial_r(r^4U^A)\pD_A\left(\frac{\e^{-2\beta}}{2r^2}\right)\\
  &-8\pi U^A\e^{-2\beta}r^2T_{rA}-8\pi r^2\e^{-2\beta}T_{ur}\,.
  \end{split}
\end{equation}
This gives us Eqs.~\eqref{eqforV} and \eqref{eqforV3} in the main text.

\section{A proof on $(\partial/\partial\phi)^\mu$ lying on a $S_r$}\label{app2}
In this appendix, we will prove following statement for stationary axisymmetric case: if (1) one orbit of vector field $\Psi^\mu$ lies on the surface $S_{r_h}$ and (2) the function $\psi$ in Eq.~\eqref{defrcoord1} satisfies $\Psi^\mu\partial_\mu\psi=0$, then any orbit of vector field $\Psi^\mu$ must lie on a $S_{r}$.

The proof contains two parts. At the first part, we prove that the orbit of $\Psi^\mu$ must lie on a null hypersurface $W_u$, i.e. $\Psi^\mu$ is tangent to null hypersurface $W_u$. To do that, we only need to prove $\Psi^\mu$ is tangent to null hypersurface $W_0$, i.e. $\Psi^\mu(\td u)_\mu|_{W_0}=0$. Because $\Psi^\mu$ is tangent to $S_{r_h}$ and $S_{r_h}\subset W_0$, we have
\begin{equation}\label{phidu01}
  \Psi^\mu(\td u)_\mu|_{S_{r_h}}=\Psi^\mu(\td u)_\mu|_{r=r_h}=0\,.
\end{equation}
Define $r^\mu=(\partial/\partial r)^\mu$. We consider the Lie derivative of $\Psi^\mu(\td u)_\mu$ with respective to $r^\mu$. The direct computation shows
\begin{equation}\label{phidu02}
\begin{split}
  \mathcal{L}_r (\Psi^\mu(\td u)_\mu)=&(\td u)_\mu\mathcal{L}_r \Psi^\mu+\Psi^\mu\mathcal{L}_r (\td u)_\mu\\
  =&(\td u)_\mu\mathcal{L}_r \Psi^\mu+\Psi^\mu(\td\mathcal{L}_r u)_\mu\,.
  \end{split}
\end{equation}
Here we use the formula $\td\mathcal{L}_r \omega=\mathcal{L}_r\td\omega$ for a $p$-form field $\omega$. As
$$\mathcal{L}_r u=r^\mu(\td u)_\mu=\psi g^{\mu\nu}(\td u)_\mu(\td u)_\nu=0\,.$$
we see
\begin{equation}\label{phidu03}
  \mathcal{L}_r (\Psi^\mu(\td u)_\mu)=(\td u)_\mu\mathcal{L}_r \Psi^\mu=-(\td u)_\mu\mathcal{L}_\Psi r^\mu\,.
\end{equation}
Using the fact that $\Psi^\mu$ is a Killing vector and the definition $r^\mu=\psi g^{\mu\nu}(\td u)_\nu$, we find
\begin{equation}\label{phidu04}
\begin{split}
  (\td u)_\mu\mathcal{L}_\Psi r^\mu=&(\td u)_\mu\mathcal{L}_\Psi [\psi g^{\mu\nu}(\td u)_\nu]\\
  =&\frac12\mathcal{L}_\Psi [(\td u)_\mu \psi g^{\mu\nu}(\td u)_\nu]=0\,.
  \end{split}
\end{equation}
Here we have used the fact $\mathcal{L}_\Psi (\psi g^{\mu\nu})=0$.
This shows that $\Psi^\mu(\td u)_\mu$ is constant along the vector $r^\mu$ and so $\Psi^\mu(\td u)_\mu=\Psi^\mu(\td u)_\mu|_{r=r_h}=0$. Thus, we find $\Psi^\mu$ is tangent to $W_0$ and all orbits of $\Psi^\mu$ lie on the surface $W_0$.

In order to prove that the orbit of $\Psi^\mu$ lies on a $S_r$, we now only need to prove $\Psi^\mu$ is tangent to $S_r$, i.e. $\Psi^\mu(\td r)_\mu=0$. The method is similar. We first note
\begin{equation}\label{liephir0}
  \Psi^\mu(\td r)_\mu|_{S_{r_h}}=0
\end{equation}
Then we consider the Lie derivative with respective to $r^\mu$
\begin{equation}\label{liephir1}
\begin{split}
  \mathcal{L}_r[\Psi^\mu(\td r)_\mu]&=\Psi^\mu(\td\mathcal{L}_r r)_\mu-(\td r)_\mu\mathcal{L}_\Psi r^\mu\\
  &=-(\td r)_\mu\mathcal{L}_\Psi r^\mu\\
  &=-(\td r)_\mu\psi g^{\mu\nu}\mathcal{L}_\Psi (\td u)_\nu\\
  &=-(\td r)_\mu\psi g^{\mu\nu}(\td \mathcal{L}_\Psi u)_\nu=0\,.
  \end{split}
\end{equation}
In the third line, it uses the facts $r^\mu=\psi g^{\mu\nu}(\td u)_\nu$ and $\mathcal{L}_\Psi(\psi g^{\mu\nu})=0$. In the last line, it uses the fact $\mathcal{L}_\Psi u=0$ as $\Psi^\mu$ is tangent to the equal-$u$ hypersurface. Eq.~\eqref{liephir1} shows that $\Psi^\mu(\td r)_\mu$ is independent or coordinate $r$. Combining this result with Eq.~\eqref{liephir0}, we find $\Psi^\mu(\td r)_\mu=0$ on the whole $W_0$. Thus, we prove that any orbit of $\Psi^\mu$ must lie on a surface $S_r$. Because of this reason, we can choose the orbit of $\Psi^\mu$ as one coordinate of $S_r$.

\bibliography{boundT-ref}

\begin{thebibliography}{33}%
\makeatletter
\providecommand \@ifxundefined [1]{%
 \@ifx{#1\undefined}
}%
\providecommand \@ifnum [1]{%
 \ifnum #1\expandafter \@firstoftwo
 \else \expandafter \@secondoftwo
 \fi
}%
\providecommand \@ifx [1]{%
 \ifx #1\expandafter \@firstoftwo
 \else \expandafter \@secondoftwo
 \fi
}%
\providecommand \natexlab [1]{#1}%
\providecommand \enquote  [1]{``#1''}%
\providecommand \bibnamefont  [1]{#1}%
\providecommand \bibfnamefont [1]{#1}%
\providecommand \citenamefont [1]{#1}%
\providecommand \href@noop [0]{\@secondoftwo}%
\providecommand \href [0]{\begingroup \@sanitize@url \@href}%
\providecommand \@href[1]{\@@startlink{#1}\@@href}%
\providecommand \@@href[1]{\endgroup#1\@@endlink}%
\providecommand \@sanitize@url [0]{\catcode `\\12\catcode `\$12\catcode
  `\&12\catcode `\#12\catcode `\^12\catcode `\_12\catcode `\%12\relax}%
\providecommand \@@startlink[1]{}%
\providecommand \@@endlink[0]{}%
\providecommand \url  [0]{\begingroup\@sanitize@url \@url }%
\providecommand \@url [1]{\endgroup\@href {#1}{\urlprefix }}%
\providecommand \urlprefix  [0]{URL }%
\providecommand \Eprint [0]{\href }%
\providecommand \doibase [0]{http://dx.doi.org/}%
\providecommand \selectlanguage [0]{\@gobble}%
\providecommand \bibinfo  [0]{\@secondoftwo}%
\providecommand \bibfield  [0]{\@secondoftwo}%
\providecommand \translation [1]{[#1]}%
\providecommand \BibitemOpen [0]{}%
\providecommand \bibitemStop [0]{}%
\providecommand \bibitemNoStop [0]{.\EOS\space}%
\providecommand \EOS [0]{\spacefactor3000\relax}%
\providecommand \BibitemShut  [1]{\csname bibitem#1\endcsname}%
\let\auto@bib@innerbib\@empty
\bibitem [{\citenamefont {Hawking}(1994)}]{Hawking:1994ss}%
  \BibitemOpen
  \bibfield  {author} {\bibinfo {author} {\bibfnamefont {S.W.}\ \bibnamefont
  {Hawking}},\ }\bibfield  {title} {\enquote {\bibinfo {title} {{Nature of
  space and time}},}\ }\href@noop {} {\  (\bibinfo {year} {1994})},\ \Eprint
  {http://arxiv.org/abs/hep-th/9409195} {arXiv:hep-th/9409195} \BibitemShut
  {NoStop}%
\bibitem [{\citenamefont {Senovilla}\ and\ \citenamefont
  {Garfinkle}(2015)}]{Senovilla:2014gza}%
  \BibitemOpen
  \bibfield  {author} {\bibinfo {author} {\bibfnamefont {Jos\'e M.~M.}\
  \bibnamefont {Senovilla}}\ and\ \bibinfo {author} {\bibfnamefont {David}\
  \bibnamefont {Garfinkle}},\ }\bibfield  {title} {\enquote {\bibinfo {title}
  {{The 1965 Penrose singularity theorem}},}\ }\href {\doibase
  10.1088/0264-9381/32/12/124008} {\bibfield  {journal} {\bibinfo  {journal}
  {Class. Quant. Grav.}\ }\textbf {\bibinfo {volume} {32}},\ \bibinfo {pages}
  {124008} (\bibinfo {year} {2015})},\ \Eprint {http://arxiv.org/abs/1410.5226}
  {arXiv:1410.5226 [gr-qc]} \BibitemShut {NoStop}%
\bibitem [{\citenamefont {Penrose}(2002)}]{Penrose2002}%
  \BibitemOpen
  \bibfield  {author} {\bibinfo {author} {\bibfnamefont {R.}~\bibnamefont
  {Penrose}},\ }\bibfield  {title} {\enquote {\bibinfo {title}
  {{{\textquotedblleft}Golden Oldie{\textquotedblright}: Gravitational
  Collapse: The Role of General Relativity}},}\ }\href {\doibase
  10.1023/a:1016578408204} {\bibfield  {journal} {\bibinfo  {journal} {General
  Relativity and Gravitation}\ }\textbf {\bibinfo {volume} {34}},\ \bibinfo
  {pages} {1141--1165} (\bibinfo {year} {2002})}\BibitemShut {NoStop}%
\bibitem [{\citenamefont {Hod}(2008)}]{Hod:2008zza}%
  \BibitemOpen
  \bibfield  {author} {\bibinfo {author} {\bibfnamefont {Shahar}\ \bibnamefont
  {Hod}},\ }\bibfield  {title} {\enquote {\bibinfo {title} {{Weak Cosmic
  Censorship: As Strong as Ever}},}\ }\href {\doibase
  10.1103/PhysRevLett.100.121101} {\bibfield  {journal} {\bibinfo  {journal}
  {Phys. Rev. Lett.}\ }\textbf {\bibinfo {volume} {100}},\ \bibinfo {pages}
  {121101} (\bibinfo {year} {2008})},\ \Eprint {http://arxiv.org/abs/0805.3873}
  {arXiv:0805.3873 [gr-qc]} \BibitemShut {NoStop}%
\bibitem [{\citenamefont {Krolak}(1986)}]{Krolak1986}%
  \BibitemOpen
  \bibfield  {author} {\bibinfo {author} {\bibfnamefont {A}~\bibnamefont
  {Krolak}},\ }\bibfield  {title} {\enquote {\bibinfo {title} {Towards the
  proof of the cosmic censorship hypothesis},}\ }\href {\doibase
  10.1088/0264-9381/3/3/004} {\bibfield  {journal} {\bibinfo  {journal}
  {Classical and Quantum Gravity}\ }\textbf {\bibinfo {volume} {3}},\ \bibinfo
  {pages} {267--280} (\bibinfo {year} {1986})}\BibitemShut {NoStop}%
\bibitem [{\citenamefont {Rangamani}(2005)}]{Rangamani2005}%
  \BibitemOpen
  \bibfield  {author} {\bibinfo {author} {\bibfnamefont {Mukund}\ \bibnamefont
  {Rangamani}},\ }\bibfield  {title} {\enquote {\bibinfo {title} {Cosmic
  censorship in {ADS}/{CFT}},}\ }in\ \href {\doibase 10.1007/1-4020-3733-3_24}
  {\emph {\bibinfo {booktitle} {String Theory: From Gauge Interactions to
  Cosmology}}}\ (\bibinfo  {publisher} {Springer Netherlands},\ \bibinfo {year}
  {2005})\ pp.\ \bibinfo {pages} {355--362}\BibitemShut {NoStop}%
\bibitem [{\citenamefont {Ong}(2020)}]{Ong:2020xwv}%
  \BibitemOpen
  \bibfield  {author} {\bibinfo {author} {\bibfnamefont {Yen~Chin}\
  \bibnamefont {Ong}},\ }\bibfield  {title} {\enquote {\bibinfo {title}
  {{Space-time singularities and cosmic censorship conjecture: A Review with
  some thoughts}},}\ }\href {\doibase 10.1142/S0217751X20300070} {\bibfield
  {journal} {\bibinfo  {journal} {Int. J. Mod. Phys.}\ }\textbf {\bibinfo
  {volume} {A35}},\ \bibinfo {pages} {2030007} (\bibinfo {year} {2020})},\
  \Eprint {http://arxiv.org/abs/2005.07032} {arXiv:2005.07032 [gr-qc]}
  \BibitemShut {NoStop}%
\bibitem [{\citenamefont {Huisken}\ and\ \citenamefont
  {Ilmanen}(2001)}]{Huisken2001}%
  \BibitemOpen
  \bibfield  {author} {\bibinfo {author} {\bibfnamefont {Gerhard}\ \bibnamefont
  {Huisken}}\ and\ \bibinfo {author} {\bibfnamefont {Tom}\ \bibnamefont
  {Ilmanen}},\ }\bibfield  {title} {\enquote {\bibinfo {title} {{The Inverse
  Mean Curvature Flow and the Riemannian Penrose Inequality}},}\ }\href
  {\doibase 10.4310/jdg/1090349447} {\bibfield  {journal} {\bibinfo  {journal}
  {Journal of Differential Geometry}\ }\textbf {\bibinfo {volume} {59}},\
  \bibinfo {pages} {353--437} (\bibinfo {year} {2001})}\BibitemShut {NoStop}%
\bibitem [{\citenamefont {Bray}(2001)}]{Bray2001}%
  \BibitemOpen
  \bibfield  {author} {\bibinfo {author} {\bibfnamefont {Hubert~L.}\
  \bibnamefont {Bray}},\ }\bibfield  {title} {\enquote {\bibinfo {title}
  {{Proof of the Riemannian Penrose Inequality Using the Positive Mass
  Theorem}},}\ }\href {\doibase 10.4310/jdg/1090349428} {\bibfield  {journal}
  {\bibinfo  {journal} {Journal of Differential Geometry}\ }\textbf {\bibinfo
  {volume} {59}},\ \bibinfo {pages} {177--267} (\bibinfo {year}
  {2001})}\BibitemShut {NoStop}%
\bibitem [{\citenamefont {Bray}\ and\ \citenamefont
  {Chrusciel}(2003)}]{Bray:2003ns}%
  \BibitemOpen
  \bibfield  {author} {\bibinfo {author} {\bibfnamefont {Hubert~L.}\
  \bibnamefont {Bray}}\ and\ \bibinfo {author} {\bibfnamefont {Piotr~T.}\
  \bibnamefont {Chrusciel}},\ }\bibfield  {title} {\enquote {\bibinfo {title}
  {{The Penrose inequality}},}\ }\href@noop {} {\  (\bibinfo {year} {2003})},\
  \Eprint {http://arxiv.org/abs/gr-qc/0312047} {arXiv:gr-qc/0312047 [gr-qc]}
  \BibitemShut {NoStop}%
\bibitem [{\citenamefont {Husain}\ and\ \citenamefont
  {Singh}(2017)}]{Husain:2017cmj}%
  \BibitemOpen
  \bibfield  {author} {\bibinfo {author} {\bibfnamefont {Viqar}\ \bibnamefont
  {Husain}}\ and\ \bibinfo {author} {\bibfnamefont {Suprit}\ \bibnamefont
  {Singh}},\ }\bibfield  {title} {\enquote {\bibinfo {title} {{Penrose
  inequality in anti\textendash{}de Sitter space}},}\ }\href {\doibase
  10.1103/PhysRevD.96.104055} {\bibfield  {journal} {\bibinfo  {journal} {Phys.
  Rev. D}\ }\textbf {\bibinfo {volume} {96}},\ \bibinfo {pages} {104055}
  (\bibinfo {year} {2017})},\ \Eprint {http://arxiv.org/abs/1709.02395}
  {arXiv:1709.02395 [gr-qc]} \BibitemShut {NoStop}%
\bibitem [{\citenamefont {Engelhardt}\ and\ \citenamefont
  {Horowitz}(2019)}]{Engelhardt:2019btp}%
  \BibitemOpen
  \bibfield  {author} {\bibinfo {author} {\bibfnamefont {Netta}\ \bibnamefont
  {Engelhardt}}\ and\ \bibinfo {author} {\bibfnamefont {Gary~T.}\ \bibnamefont
  {Horowitz}},\ }\bibfield  {title} {\enquote {\bibinfo {title} {{Holographic
  argument for the Penrose inequality in AdS spacetimes}},}\ }\href {\doibase
  10.1103/PhysRevD.99.126009} {\bibfield  {journal} {\bibinfo  {journal} {Phys.
  Rev. D}\ }\textbf {\bibinfo {volume} {99}},\ \bibinfo {pages} {126009}
  (\bibinfo {year} {2019})},\ \Eprint {http://arxiv.org/abs/1903.00555}
  {arXiv:1903.00555 [hep-th]} \BibitemShut {NoStop}%
\bibitem [{\citenamefont {Bousso}\ \emph
  {et~al.}(2019{\natexlab{a}})\citenamefont {Bousso}, \citenamefont
  {Shahbazi-Moghaddam},\ and\ \citenamefont {Tomasevic}}]{Bousso:2019var}%
  \BibitemOpen
  \bibfield  {author} {\bibinfo {author} {\bibfnamefont {Raphael}\ \bibnamefont
  {Bousso}}, \bibinfo {author} {\bibfnamefont {Arvin}\ \bibnamefont
  {Shahbazi-Moghaddam}}, \ and\ \bibinfo {author} {\bibfnamefont {Marija}\
  \bibnamefont {Tomasevic}},\ }\bibfield  {title} {\enquote {\bibinfo {title}
  {{Quantum Penrose Inequality}},}\ }\href {\doibase
  10.1103/PhysRevLett.123.241301} {\bibfield  {journal} {\bibinfo  {journal}
  {Phys. Rev. Lett.}\ }\textbf {\bibinfo {volume} {123}},\ \bibinfo {pages}
  {241301} (\bibinfo {year} {2019}{\natexlab{a}})},\ \Eprint
  {http://arxiv.org/abs/1908.02755} {arXiv:1908.02755 [hep-th]} \BibitemShut
  {NoStop}%
\bibitem [{\citenamefont {Bousso}\ \emph
  {et~al.}(2019{\natexlab{b}})\citenamefont {Bousso}, \citenamefont
  {Shahbazi-Moghaddam},\ and\ \citenamefont
  {Toma\v{s}evi\'c}}]{Bousso:2019bkg}%
  \BibitemOpen
  \bibfield  {author} {\bibinfo {author} {\bibfnamefont {Raphael}\ \bibnamefont
  {Bousso}}, \bibinfo {author} {\bibfnamefont {Arvin}\ \bibnamefont
  {Shahbazi-Moghaddam}}, \ and\ \bibinfo {author} {\bibfnamefont {Marija}\
  \bibnamefont {Toma\v{s}evi\'c}},\ }\bibfield  {title} {\enquote {\bibinfo
  {title} {{Quantum Information Bound on the Energy}},}\ }\href {\doibase
  10.1103/PhysRevD.100.126010} {\bibfield  {journal} {\bibinfo  {journal}
  {Phys. Rev. D}\ }\textbf {\bibinfo {volume} {100}},\ \bibinfo {pages}
  {126010} (\bibinfo {year} {2019}{\natexlab{b}})},\ \Eprint
  {http://arxiv.org/abs/1909.02001} {arXiv:1909.02001 [hep-th]} \BibitemShut
  {NoStop}%
\bibitem [{\citenamefont {Brown}\ \emph {et~al.}(1994)\citenamefont {Brown},
  \citenamefont {Creighton},\ and\ \citenamefont {Mann}}]{PhysRevD.50.6394}%
  \BibitemOpen
  \bibfield  {author} {\bibinfo {author} {\bibfnamefont {J.~D.}\ \bibnamefont
  {Brown}}, \bibinfo {author} {\bibfnamefont {J.}~\bibnamefont {Creighton}}, \
  and\ \bibinfo {author} {\bibfnamefont {R.~B.}\ \bibnamefont {Mann}},\
  }\bibfield  {title} {\enquote {\bibinfo {title} {Temperature, energy, and
  heat capacity of asymptotically anti--de sitter black holes},}\ }\href
  {\doibase 10.1103/PhysRevD.50.6394} {\bibfield  {journal} {\bibinfo
  {journal} {Phys. Rev. D}\ }\textbf {\bibinfo {volume} {50}},\ \bibinfo
  {pages} {6394--6403} (\bibinfo {year} {1994})}\BibitemShut {NoStop}%
\bibitem [{\citenamefont {Comer}(1992)}]{Comer1992}%
  \BibitemOpen
  \bibfield  {author} {\bibinfo {author} {\bibfnamefont {G~L}\ \bibnamefont
  {Comer}},\ }\bibfield  {title} {\enquote {\bibinfo {title} {Ensemble
  dependence of the stability of thermal black holes},}\ }\href {\doibase
  10.1088/0264-9381/9/4/011} {\bibfield  {journal} {\bibinfo  {journal}
  {Classical and Quantum Gravity}\ }\textbf {\bibinfo {volume} {9}},\ \bibinfo
  {pages} {947--962} (\bibinfo {year} {1992})}\BibitemShut {NoStop}%
\bibitem [{\citenamefont {Quevedo}\ \emph {et~al.}(2014)\citenamefont
  {Quevedo}, \citenamefont {Quevedo}, \citenamefont {Sanchez},\ and\
  \citenamefont {Taj}}]{Quevedo:2013pba}%
  \BibitemOpen
  \bibfield  {author} {\bibinfo {author} {\bibfnamefont {Hernando}\
  \bibnamefont {Quevedo}}, \bibinfo {author} {\bibfnamefont {Maria~N.}\
  \bibnamefont {Quevedo}}, \bibinfo {author} {\bibfnamefont {Alberto}\
  \bibnamefont {Sanchez}}, \ and\ \bibinfo {author} {\bibfnamefont {Safia}\
  \bibnamefont {Taj}},\ }\bibfield  {title} {\enquote {\bibinfo {title} {{On
  the ensemble dependence in black hole geometrothermodynamics}},}\ }\href
  {\doibase 10.1088/0031-8949/89/8/084007} {\bibfield  {journal} {\bibinfo
  {journal} {Phys. Scripta}\ }\textbf {\bibinfo {volume} {89}},\ \bibinfo
  {pages} {084007} (\bibinfo {year} {2014})},\ \Eprint
  {http://arxiv.org/abs/1304.3954} {arXiv:1304.3954 [gr-qc]} \BibitemShut
  {NoStop}%
\bibitem [{\citenamefont {Visser}(1992)}]{Visser:1992qh}%
  \BibitemOpen
  \bibfield  {author} {\bibinfo {author} {\bibfnamefont {Matt}\ \bibnamefont
  {Visser}},\ }\bibfield  {title} {\enquote {\bibinfo {title} {{Dirty black
  holes: Thermodynamics and horizon structure}},}\ }\href {\doibase
  10.1103/PhysRevD.46.2445} {\bibfield  {journal} {\bibinfo  {journal} {Phys.
  Rev. D}\ }\textbf {\bibinfo {volume} {46}},\ \bibinfo {pages} {2445--2451}
  (\bibinfo {year} {1992})},\ \Eprint {http://arxiv.org/abs/hep-th/9203057}
  {arXiv:hep-th/9203057} \BibitemShut {NoStop}%
\bibitem [{\citenamefont {Carter}(1971)}]{PhysRevLett.26.331}%
  \BibitemOpen
  \bibfield  {author} {\bibinfo {author} {\bibfnamefont {B.}~\bibnamefont
  {Carter}},\ }\bibfield  {title} {\enquote {\bibinfo {title} {Axisymmetric
  black hole has only two degrees of freedom},}\ }\href {\doibase
  10.1103/PhysRevLett.26.331} {\bibfield  {journal} {\bibinfo  {journal} {Phys.
  Rev. Lett.}\ }\textbf {\bibinfo {volume} {26}},\ \bibinfo {pages} {331--333}
  (\bibinfo {year} {1971})}\BibitemShut {NoStop}%
\bibitem [{\citenamefont {Hawking}(1972)}]{hawking1972}%
  \BibitemOpen
  \bibfield  {author} {\bibinfo {author} {\bibfnamefont {S.~W.}\ \bibnamefont
  {Hawking}},\ }\bibfield  {title} {\enquote {\bibinfo {title} {Black holes in
  general relativity},}\ }\href
  {https://projecteuclid.org:443/euclid.cmp/1103857884} {\bibfield  {journal}
  {\bibinfo  {journal} {Comm. Math. Phys.}\ }\textbf {\bibinfo {volume} {25}},\
  \bibinfo {pages} {152--166} (\bibinfo {year} {1972})}\BibitemShut {NoStop}%
\bibitem [{\citenamefont {Hawking}\ and\ \citenamefont
  {Ellis}(2011)}]{Hawking:1973uf}%
  \BibitemOpen
  \bibfield  {author} {\bibinfo {author} {\bibfnamefont {S.W.}\ \bibnamefont
  {Hawking}}\ and\ \bibinfo {author} {\bibfnamefont {G.F.R.}\ \bibnamefont
  {Ellis}},\ }\href {\doibase 10.1017/CBO9780511524646} {\emph {\bibinfo
  {title} {{The Large Scale Structure of Space-Time}}}},\ Cambridge Monographs
  on Mathematical Physics\ (\bibinfo  {publisher} {Cambridge University
  Press},\ \bibinfo {year} {2011})\BibitemShut {NoStop}%
\bibitem [{\citenamefont {Wald}(1984)}]{robertwald1984}%
  \BibitemOpen
  \bibfield  {author} {\bibinfo {author} {\bibfnamefont {Robert~M.}\
  \bibnamefont {Wald}},\ }\href {https://www.xarg.org/ref/a/0226870332/} {\emph
  {\bibinfo {title} {General Relativity}}}\ (\bibinfo  {publisher} {University
  of Chicago Press},\ \bibinfo {year} {1984})\BibitemShut {NoStop}%
\bibitem [{\citenamefont {Galloway}\ and\ \citenamefont
  {Schoen}(2006)}]{Galloway:2005mf}%
  \BibitemOpen
  \bibfield  {author} {\bibinfo {author} {\bibfnamefont {Gregory~J.}\
  \bibnamefont {Galloway}}\ and\ \bibinfo {author} {\bibfnamefont {Richard}\
  \bibnamefont {Schoen}},\ }\bibfield  {title} {\enquote {\bibinfo {title} {{A
  Generalization of Hawking's black hole topology theorem to higher
  dimensions}},}\ }\href {\doibase 10.1007/s00220-006-0019-z} {\bibfield
  {journal} {\bibinfo  {journal} {Commun. Math. Phys.}\ }\textbf {\bibinfo
  {volume} {266}},\ \bibinfo {pages} {571--576} (\bibinfo {year} {2006})},\
  \Eprint {http://arxiv.org/abs/gr-qc/0509107} {arXiv:gr-qc/0509107}
  \BibitemShut {NoStop}%
\bibitem [{\citenamefont {Racz}\ and\ \citenamefont
  {Wald}(1996)}]{Racz:1995nh}%
  \BibitemOpen
  \bibfield  {author} {\bibinfo {author} {\bibfnamefont {Istvan}\ \bibnamefont
  {Racz}}\ and\ \bibinfo {author} {\bibfnamefont {Robert~M.}\ \bibnamefont
  {Wald}},\ }\bibfield  {title} {\enquote {\bibinfo {title} {{Global extensions
  of space-times describing asymptotic final states of black holes}},}\ }\href
  {\doibase 10.1088/0264-9381/13/3/017} {\bibfield  {journal} {\bibinfo
  {journal} {Class. Quant. Grav.}\ }\textbf {\bibinfo {volume} {13}},\ \bibinfo
  {pages} {539--553} (\bibinfo {year} {1996})},\ \Eprint
  {http://arxiv.org/abs/gr-qc/9507055} {arXiv:gr-qc/9507055} \BibitemShut
  {NoStop}%
\bibitem [{\citenamefont {Hawking}(1971)}]{Hawking1971}%
  \BibitemOpen
  \bibfield  {author} {\bibinfo {author} {\bibfnamefont {S.~W.}\ \bibnamefont
  {Hawking}},\ }\bibfield  {title} {\enquote {\bibinfo {title} {Gravitational
  radiation from colliding black holes},}\ }\href {\doibase
  10.1103/physrevlett.26.1344} {\bibfield  {journal} {\bibinfo  {journal}
  {Physical Review Letters}\ }\textbf {\bibinfo {volume} {26}},\ \bibinfo
  {pages} {1344--1346} (\bibinfo {year} {1971})}\BibitemShut {NoStop}%
\bibitem [{\citenamefont {Hong}\ \emph
  {et~al.}(2020{\natexlab{a}})\citenamefont {Hong}, \citenamefont {Suzuki},\
  and\ \citenamefont {Yamada}}]{Hong:2019mcj}%
  \BibitemOpen
  \bibfield  {author} {\bibinfo {author} {\bibfnamefont {Jeong-Pyong}\
  \bibnamefont {Hong}}, \bibinfo {author} {\bibfnamefont {Motoo}\ \bibnamefont
  {Suzuki}}, \ and\ \bibinfo {author} {\bibfnamefont {Masaki}\ \bibnamefont
  {Yamada}},\ }\bibfield  {title} {\enquote {\bibinfo {title} {{Charged black
  holes in non-linear Q-clouds with O(3) symmetry}},}\ }\href {\doibase
  10.1016/j.physletb.2020.135324} {\bibfield  {journal} {\bibinfo  {journal}
  {Phys. Lett. B}\ }\textbf {\bibinfo {volume} {803}},\ \bibinfo {pages}
  {135324} (\bibinfo {year} {2020}{\natexlab{a}})},\ \Eprint
  {http://arxiv.org/abs/1907.04982} {arXiv:1907.04982 [gr-qc]} \BibitemShut
  {NoStop}%
\bibitem [{\citenamefont {Hong}\ \emph
  {et~al.}(2020{\natexlab{b}})\citenamefont {Hong}, \citenamefont {Suzuki},\
  and\ \citenamefont {Yamada}}]{Hong:2020miv}%
  \BibitemOpen
  \bibfield  {author} {\bibinfo {author} {\bibfnamefont {Jeong-Pyong}\
  \bibnamefont {Hong}}, \bibinfo {author} {\bibfnamefont {Motoo}\ \bibnamefont
  {Suzuki}}, \ and\ \bibinfo {author} {\bibfnamefont {Masaki}\ \bibnamefont
  {Yamada}},\ }\bibfield  {title} {\enquote {\bibinfo {title} {{Q-hairs and
  no-hair theorem for charged black holes}},}\ }\href {\doibase
  10.1103/PhysRevLett.125.111104} {\bibfield  {journal} {\bibinfo  {journal}
  {Phys. Rev. Lett.}\ }\textbf {\bibinfo {volume} {125}},\ \bibinfo {pages}
  {111104} (\bibinfo {year} {2020}{\natexlab{b}})},\ \Eprint
  {http://arxiv.org/abs/2004.03148} {arXiv:2004.03148 [gr-qc]} \BibitemShut
  {NoStop}%
\bibitem [{\citenamefont {BONDI}(1960)}]{BONDI1960}%
  \BibitemOpen
  \bibfield  {author} {\bibinfo {author} {\bibfnamefont {H.}~\bibnamefont
  {BONDI}},\ }\bibfield  {title} {\enquote {\bibinfo {title} {Gravitational
  waves in general relativity},}\ }\href {\doibase 10.1038/186535a0} {\bibfield
   {journal} {\bibinfo  {journal} {Nature}\ }\textbf {\bibinfo {volume}
  {186}},\ \bibinfo {pages} {535--535} (\bibinfo {year} {1960})}\BibitemShut
  {NoStop}%
\bibitem [{\citenamefont {Sachs}(1962)}]{Sachs}%
  \BibitemOpen
  \bibfield  {author} {\bibinfo {author} {\bibfnamefont {R.~K.}\ \bibnamefont
  {Sachs}},\ }\bibfield  {title} {\enquote {\bibinfo {title} {Gravitational
  waves in general relativity {VIII}. waves in asymptotically flat
  space-time},}\ }\href {\doibase 10.1098/rspa.1962.0206} {\bibfield  {journal}
  {\bibinfo  {journal} {Proceedings of the Royal Society of London. Series A.
  Mathematical and Physical Sciences}\ }\textbf {\bibinfo {volume} {270}},\
  \bibinfo {pages} {103--126} (\bibinfo {year} {1962})}\BibitemShut {NoStop}%
\bibitem [{\citenamefont {Cao}\ and\ \citenamefont {He}(2013)}]{Cao2013}%
  \BibitemOpen
  \bibfield  {author} {\bibinfo {author} {\bibfnamefont {Zhoujian}\
  \bibnamefont {Cao}}\ and\ \bibinfo {author} {\bibfnamefont {Xiaokai}\
  \bibnamefont {He}},\ }\bibfield  {title} {\enquote {\bibinfo {title}
  {Generalized bondi-sachs equations for characteristic formalism of numerical
  relativity},}\ }\href {\doibase 10.1103/physrevd.88.104002} {\bibfield
  {journal} {\bibinfo  {journal} {Physical Review D}\ }\textbf {\bibinfo
  {volume} {88}} (\bibinfo {year} {2013}),\
  10.1103/physrevd.88.104002}\BibitemShut {NoStop}%
\bibitem [{\citenamefont {Winicour}(1983)}]{doi:10.1063/1.525796}%
  \BibitemOpen
  \bibfield  {author} {\bibinfo {author} {\bibfnamefont {J.}~\bibnamefont
  {Winicour}},\ }\bibfield  {title} {\enquote {\bibinfo {title} {Newtonian
  gravity on the null cone},}\ }\href {\doibase 10.1063/1.525796} {\bibfield
  {journal} {\bibinfo  {journal} {Journal of Mathematical Physics}\ }\textbf
  {\bibinfo {volume} {24}},\ \bibinfo {pages} {1193--1198} (\bibinfo {year}
  {1983})},\ \Eprint {http://arxiv.org/abs/https://doi.org/10.1063/1.525796}
  {https://doi.org/10.1063/1.525796} \BibitemShut {NoStop}%
\bibitem [{\citenamefont {M\"{a}dler}\ and\ \citenamefont
  {Winicour}(2016)}]{M_dler_2016}%
  \BibitemOpen
  \bibfield  {author} {\bibinfo {author} {\bibfnamefont {Thomas}\ \bibnamefont
  {M\"{a}dler}}\ and\ \bibinfo {author} {\bibfnamefont {Jeffrey}\ \bibnamefont
  {Winicour}},\ }\bibfield  {title} {\enquote {\bibinfo {title} {Bondi-sachs
  formalism},}\ }\href {\doibase 10.4249/scholarpedia.33528} {\bibfield
  {journal} {\bibinfo  {journal} {Scholarpedia}\ }\textbf {\bibinfo {volume}
  {11}},\ \bibinfo {pages} {33528} (\bibinfo {year} {2016})}\BibitemShut
  {NoStop}%
\bibitem [{\citenamefont {Wald}(2001)}]{Wald:1999vt}%
  \BibitemOpen
  \bibfield  {author} {\bibinfo {author} {\bibfnamefont {Robert~M.}\
  \bibnamefont {Wald}},\ }\bibfield  {title} {\enquote {\bibinfo {title} {{The
  thermodynamics of black holes}},}\ }\href {\doibase 10.12942/lrr-2001-6}
  {\bibfield  {journal} {\bibinfo  {journal} {Living Rev. Rel.}\ }\textbf
  {\bibinfo {volume} {4}},\ \bibinfo {pages} {6} (\bibinfo {year} {2001})},\
  \Eprint {http://arxiv.org/abs/gr-qc/9912119} {arXiv:gr-qc/9912119}
  \BibitemShut {NoStop}%
\end{thebibliography}%

\end{document}